\documentclass[twoside,11pt]{article}
\usepackage{amsmath}
\usepackage{algorithm}
\usepackage{algpseudocode}
\usepackage{float, multirow,parskip, subcaption, setspace}
\usepackage{graphicx}
\usepackage{comment}
\usepackage{url}
\usepackage{enumitem}
\usepackage{tikz}
\usepackage{bm, bbm}
\usetikzlibrary{shapes.geometric, positioning}
\usetikzlibrary{quotes, angles}
\usepackage{rotating}
\usepackage{xr, xr-hyper}
\usepackage{hyperref}

\definecolor{SkyBlue}{RGB}{14, 118, 188}
\definecolor{BrightRed}{RGB}{223, 82, 78}
\definecolor{Green638}{RGB}{165,255,118} 

\hypersetup{pdfborder = {0 0 0.5 [3 3]}, colorlinks = true, linkcolor = BrightRed, citecolor = SkyBlue, filecolor = BrightRed}

\newcommand{\Rlang}{\textsf{R}}

\newcommand{\R}{\mathbb{R}} 
\newcommand{\E}{\mathbb{E}} 
\def\P{\mathbb{P}} 

\newcommand{\catt}{\textrm{CATT}}

\newcommand{\ind}[1]{\mathbbm{1}\left( #1 \right)} 

\newcommand{\bx}{\bm{x}}

\newcommand{\bX}{\bm{X}}

\newcommand{\indep}{\perp\kern-0.6em\perp}

\usepackage{amssymb, amsthm, fullpage, natbib}

\title{Bayesian Causal Forests \& the 2022 ACIC Data Challenge: Scalability and Sensitivity}
\author{Ajinkya H. Kokandakar\thanks{Dept.~of Statistics, University of Wisconsin--Madison. Correspondence to: \nolinkurl{sameer.deshpande@wisc.edu}},\  Hyunseung Kang\footnotemark[1], and Sameer K. Deshpande\footnotemark[1]}

\begin{document}

\maketitle

\begin{abstract}
We demonstrate how Hahn et al.'s Bayesian Causal Forests model (BCF) can be used to estimate conditional average treatment effects for the longitudinal dataset in the 2022 American Causal Inference Conference Data Challenge. 
Unfortunately, existing implementations of BCF do not scale to the size of the challenge data.
Therefore, we developed \textbf{flexBCF} --- a more scalable and flexible implementation of BCF --- and used it in our challenge submission. 
We investigate the sensitivity of our results to the choice of propensity score estimation method and the use of sparsity-inducing regression tree priors.
While we found that our overall point predictions were not especially sensitive to these modeling choices, we did observe that running BCF with flexibly estimated propensity scores often yielded better-calibrated uncertainty intervals. 
\end{abstract}

\newpage
\section{Introduction}
\label{sec:introduction}
Given pairs of observed predictor vectors $\bx \in \R^{p}$ and scalar outputs $y \in \R,$ \citet{Chipman2010}'s Bayesian additive regression trees (BART) model approximates the regression function $f(\bx) := \E[y \vert \bx]$ with a sum of binary regression trees.
BART often delivers accurate point estimates and well-calibrated uncertainty intervals around evaluations $f(\bx)$ without requiring users to (i) pre-specify the functional form of $f$ or (ii) tune any hyperparameters.
BART's ease-of-use and generally excellent tuning-free estimation, prediction, and uncertainty quantification have made BART a popular ``off-the-shelf'' modeling tool.

As \citet{Hill2011Bayesian} noted, these advantages make BART an attractive choice for estimating heterogeneous treatment effects.
To wit, conditional average treatment effects may be non-linear and may depend on complicated interactions between confounders, covariates, and treatment.
Correctly specifying such non-linearities and interactions is difficult --- if not impossible --- in most practical settings, making BART particularly compelling.
Given observed triplets $(\bx, y, z)$ of covariates $\bx$, outcome $y$, and treatment indicator $z,$ \citet{Hill2011Bayesian} used BART to obtain an estimate $\hat{f}(\bx, z)$ of the response surface $\E[Y \vert \bX = \bx, Z = z]$ before estimating treatment effects as $\hat{f}(\bx,1) - \hat{f}(\bx,0).$ 
In past iterations of the ACIC Data Challenge, BART-based models have consistently ranked among the top-performing methods.
One method in particular --- \citet{Hahn2020}'s Bayesian causal forests (BCF) model --- stands out as particularly adept at estimating heterogeneous treatment effects.
In this paper, we describe our experience deploying BCF to the 2022 ACIC Data Challenge.

At a high-level, BCF models the expected outcome as $\E[Y \vert \bX = \bx, Z = z] = \mu(\bx) + z \times \tau(\bx)$
and further approximates the prognostic function $\mu$ and the treatment effect function $\tau$ using separate regression tree ensembles.
When fitting the tree ensembles, BCF includes an estimate of the propensity score as an additional covariate.
Although these ensembles are \textit{a priori} independent, they are dependent \textit{a posteriori}, as BCF leverages all observations (both treated and control) to learn $\mu.$

At first glance, BCF appears ill-suited to the 2022 ACIC Data Challenge: BCF was initially proposed for cross-sectional data and the Data Challenge features longitudinal data with time-varying covariates and outcomes.
However, as we show in Section~\ref{sec:identification}, under a mild assumption about the data generating process, the Data Challenge admits an additive decomposition similar to the one used by BCF.
Unfortunately, we found that the implementation available in the \Rlang~package \textbf{bcf} did not scale to the challenge data.
We therefore re-implemented a version of BCF that can scale to the Data Challenge.

When writing our implementation and subsequently deploying it during the competition, we made several essentially \textit{ad hoc} modeling decisions.
First, we included an estimate of \citet{ImaiRatkovic2014}'s covariate balancing propensity score (CBPS) as an additional covariate in the expected outcome model. 
Second, in our regression tree prior, we selected splitting variables using \citet{Linero2018}'s sparsity-inducing Dirichlet prior instead of the uniform prior that is used in the \textbf{bcf} package. 
Briefly, \citet{Linero2018}'s prior encourages regression trees to split on a small number of covariates, thereby facilitating a type of automatic variable selection.
In the context of treatment effect estimation, such a sparsity-inducing prior can potentially identify the main drivers of effect heterogeneity.
We assess the sensitivity of our contest submissions to these modeling choices. 

Here is an outline for the rest of the paper.
We briefly introduce the competition dataset in Section~\ref{sec:methods} before presenting our identification analysis, introducing our new implementation of BCF, and reporting the results of our initial challenge submission.
Then, in Section~\ref{sec:results}, we describe our post-challenge sensitivity analysis.
Finally, we discuss the implications of our findings in Section~\ref{sec:discussion}.

\section{Identification and estimation}
\label{sec:methods}
\subsection{Review: Study design, notation, and definitions}
\label{sec:data_description}
The 2022 ACIC Data Challenge involved analyzing a total of 3,400 synthetic datasets consisting of 200 independent realizations of 17 different data generating processes (DGP's).
Each synthetic dataset simulated observations of beneficiary-level Medicare expenditure data over a period of four years.
Each beneficiary received care at one of about 500 different medical practices. 
In addition to simulated time-invariant beneficiary- and practice-level covariates, each dataset also included average measurements of patient covariates within each practice.
In each DGP, treatment was assigned to some practices after two years.
The main task of the Challenge was to assess how the treatment deployed at the practice level affected future medical expenditures of beneficiaries in the treated practices. 

For each beneficiary $i$ in practice $j$, let $Y_{ijt}$ denote their outcome in year $t$. 
Also, for each practice $j$, let $Z_{j} \in \{0,1\}$ denote its treatment status where  $Z_{j} = 1$ indicates that practice $j$ was assigned treatment between year 2 and year 3 and let $Z_{j} = 0$ indicates that practice $j$ was not treated during that period. 
As the notation suggests, the treatment is assigned only once and before assigning treatment, all practices are untreated. 
Also, treatment is assigned at the practice level and all beneficiaries in a practice received the same treatment condition. 
Hence, we omit subscripts $i$ and $t$ in $Z_{j}$ for notational simplicity.
Finally, for each beneficiary $i$ in practice $j$, let $\bX_{ij} \in \mathbb{R}^p$ be all observed pre-treatment covariates, i.e.~non-outcome variables measured at years $1$ and  $2$ before treatment is assigned. 
To prevent potential feedback, when modeling the full set of observed outcomes, we did not include $Y_{ij1}$ or $Y_{ij2}$ in $\bX_{ij}.$
Again, for notational simplicity, we omit the subscript $t$ in $\bX_{ij}.$

We define causal effects using potential outcomes \citep{Neyman1923, Rubin1974}.
For each beneficiary $i$ in practice $j$, let $Y_{ijt}(0)$ denote their potential outcome at time $t$ when their practice was untreated between years 2 and 3.
Similarly, let $Y_{ijt}(1)$ denote their potential outcome at time $t$ when their practice was treated between years 2 and 3.
Implicit in our notation are the assumptions that (i) the treatment status of practice $j$ cannot affect the potential outcome of beneficiaries in practice $j' \neq j$ and (ii) there is only one version of treatment and only one version of non-treatment.
Together these assumptions are often know as the Stable Unit Treatment Value Assumption \citep[SUTVA]{Rubin1980_randomization, Rubin1986_comment}.
We consider these assumptions reasonable in the context of the Data Challenge, while also noting the Challenge provided insufficient context about the data to probe potential violations\footnote{In fact, our notation makes further implicit assumptions that we similarly regard as reasonable. Namely, we assume that treatment status does not vary amongst beneficiaries in the same practice and does not change once assigned in between years 2 and 3.}.

We focus on the conditional average treatment effect on the treated (CATT), defined as
\begin{equation}\label{eq:estimand}
    \catt(\bx, t) = \mathbb{E}\left[Y_{ijt}(1)-Y_{ijt}(0) \vert Z_j =  1, \bX_{ij} = \bx\right], \quad{} t \in \{3,4\}.
\end{equation}
Averaging the $\catt$ with respect to the distribution of covariates yields the average treatment effect on the treated (ATT), i.e.~${\rm ATT}(t) = \mathbb{E}\left[\catt(\bX_{ij}, t)\right]$. 

\subsection{Review: Causal Identification}
\label{sec:identification}

To identify the $\catt$ from the observed data, we made the following three assumptions for all beneficiaries and practices:
\begin{itemize}
\item[(A1)]{For all $t$, $Y_{ijt} =Y_{ijt}(0) \times \ind{t \leq 2} +  Y_{ijt}(Z_j) \times \ind{t > 2}$}
\item[(A2)]{For each $s \in \{1,2\}$ and $t \in \{3,4\},$ $(Y_{ijt}(0) - Y_{ijs}(0)) \indep Z_{j} \mid \bX_{ij}$.}
\item[(A3)]{There is a $0 < \delta < 1$ such that $\delta < \P(Z_{j} = 1 \vert \bm{X}_{ij} = \bm{x}) < 1-\delta$ for all $\bx$.}
\end{itemize}
As written, (A1) implicitly assumes that (i) the observed outcomes of beneficiaries in practice $j$ does not depend on the treatment status of a different practice $j'$; and that (ii) the observed outcome of beneficiary $i$ in practice $j$ does not depend on the potential outcomes of any other beneficiaries.
Because treatment is defined at the practice level and not the beneficiary level, to paraphrase \citet[\S 1.2]{Small2008_group}, the issue of interference between beneficiaries in the same practice does not arise by design.
Further, without additional context about the motivating application, it is impossible to probe potential violations of these non-interference assumptions with the provided data.
We further note that (A3) is somewhat stronger than the assumption imposed by the Data Challenge organizers.
Namely, our assumption applies to all practices and not just the treated practices.

Assuming (A1)--(A3), we can write $\catt(\bx, t)$ as a function of the observed data:
\begin{align}
\label{eq:main_identification}
    \begin{split}
    \catt(\bx, t) \ =\ & 
        \mathbb{E}\left[Y_{ijt} \mid Z_j = 1, \bX_{ij} = \bx \right] - \mathbb{E}\left[Y_{ijt} \mid Z_j = 0, \bX_{ij} = \bx \right]\\ 
        &- \Big\{ 
        \mathbb{E}\left[Y_{ijs} \mid Z_j = 1, \bX_{ij} = \bx \right] - \mathbb{E}\left[Y_{ijs} \mid Z_j = 0, \bX_{ij } = \bx \right]\Big\}. 
    \end{split}
\end{align}
See Appendix~\ref{app:identification} for a proof.

Notice that, according to Equation~\eqref{eq:main_identification}, we can identify the $\catt$ using either $s = 1$ or $s = 2.$
This introduces a further constraint, namely that
\begin{align} \label{eq:eq1}
\mathbb{E}\left[Y_{ij2} - Y_{ij1} \mid Z_{j} = 1, \bX_{ij } = \bx \right] &=  \mathbb{E}\left[Y_{ij2} - Y_{ij1} \mid Z_{j} = 0, \bX_{ij } = \bx \right].  
\end{align}
In words, Equation \eqref{eq:eq1} roughly states that conditional on covariates, the average pre-treatment trend in outcomes is identical between the treated and the untreated groups. 
The additional structure implied by these assumptions forms the basis for an additional modeling assumption that eventually allows us to deploy BCF.

\subsection{Estimation \& Implementation}
\label{sec:estimation}

We will additionally assume that for every $(t, \bx, z)$, we can write
\begin{equation}\label{eq:model}
    \E\left[Y_{ijt} \vert \bX_{ij} = \bx, Z_j = z \right] = \mu(\bx, t)  + z \times \ind{t > 2} \times \tau(\bx, t).
\end{equation}
Under Equation~\eqref{eq:model} and the assumptions introduced in Section~\ref{sec:identification}, $\catt(\bx,t) = \tau(\bx,t).$ 
Furthermore, $\mathbb{E}\left[Y_{ij2} - Y_{ij1} \vert Z_{j} = z, \bX_{ij } = \bx \right] = \mu(\bx,2)-\mu(\bx,1)$ for $z \in \{0,1\}$, satisfying Equation~\eqref{eq:eq1}. 
In contrast to Assumption (A2), which was about trends in outcomes, Equation~\eqref{eq:model} makes an assumption directly about the outcome at each time $t.$
We note that the functional form of the expression on the right-hand side of Equation~\eqref{eq:model} is very similar to that used by BCF.

On this basis, we propose fitting the model in Equation~\eqref{eq:model} with BCF.
That is, we will approximate each of $\mu$ and $\tau$ with their own ensemble of regression trees, which receive their own independent priors.
As noted by \citet{Hahn2020}, using an additive decomposition as in Equation~\eqref{eq:model} allows us to regularize our estimate of $\tau$ in a direct and transparent fashion.
This is in sharp contrast to approaches like \citet{Hill2011Bayesian} that flexibly fit the response surface $\E[Y_{ijt} \vert \bX_{ij}, Z_{j}]$ with a single regression tree ensemble.

\subsection{Introducing \textbf{flexBCF}}
\label{sec:flexBCF}

Unfortunately, we encountered several problems when we tried to use the \textbf{bcf} package to fit the beneficiary-level data. 
We discovered that \textbf{bcf} makes a copy of all covariate data and then allocates a large matrix to store posterior samples of treatment effect function evaluations.
When modeling the beneficiary-level data, we included the practice label as an additional categorical covariate, which \textbf{bcf} subsequently converted into binary indicators (one per practice).
As a result, \textbf{bcf} allocated 5GB of memory to copy the covariate data.
\textbf{bcf} further attempted to allocate 20GB to store 2000 posterior samples (the number used in the examples in \textbf{bcf}'s documentation) for every $\tau(\bx_{ij},t)$ value.
In this way, before it could even begin the main MCMC simulation, \textbf{bcf}'s memory requirement far exceeded the capacity of our personal computers.

When we tried running \textbf{bcf} on a high-memory computer, we found that the sampler was extremely slow. 
On further inspection, we discovered that the posterior updates of each regression tree performed many redundant computations. 
Basically, while updating a single tree, \textbf{bcf} loops over the entire dataset several times to identify the leaf to which every observation is assigned.
Because the underlying Gibbs sampler only changes at most two leaves at a time, the map from observations to tree leaves does not change much iteration to iteration, rendering many of these loops redundant.

To overcome these difficulties, we developed the \Rlang~package \textbf{flexBCF}, which is available at \url{https://github.com/skdeshpande91/flexBCF}.
\textbf{flexBCF} includes a more efficient implementation of the basic BCF model that does not (i) unnecessarily copy the data; (ii) perform redundant calculations when updating the trees; and (iii) instantiate a matrix to hold all posterior draws of $\tau(\bx,t).$
Instead, our sampler returns a list of character vectors whose elements contain string representations of each regression tree.
\textbf{flexBCF} includes a predict routine that takes in these character vectors, a new set of covariates $\bx,$ and a time index $t,$ and returns posterior samples of $\tau(\bx,t).$
As a result, \textbf{flexBCF} is faster and less memory-intensive than \textbf{bcf}: for each beneficiary-level dataset, it only took a few hours and about 4GB of memory to run multiple MCMC chains in sequence. 

\textbf{flexBCF} also introduces some methodological changes aimed at achieving more flexible fits to data.
Namely, our package gives users the option to use \citet{Linero2018}'s sparsity-inducing prior on the splitting variables used in the $\mu$ and $\tau$ ensembles.
More substantively, unlike \textbf{bcf} and most implementations of BART-based models, \textbf{flexBCF} does not one-hot encode categorical predictors.
In the context of the Data Challenge, such one-hot encoding forces \textbf{bcf} to partition the practices using a recursive ``remove one at a time'' strategy.
This strategy partitions the practices into a few singleton sets and one large set with the remaining practices.
In this way, \textbf{bcf} is prevented from forming several small subgroups of practices and partially pooling data within each group.
By directly partitioning the practices instead of binary indicators, \textbf{flexBCF} can form a richer set of partitions, endowing it with much more representational flexibility.
See \citet{Deshpande2022_flexBART} for an argument against one-hot encoding categorical predictors in BART-based models and for a description of an alternative strategy, which is employed by \textbf{flexBCF}.

\subsection{Our initial challenge submission}
\label{sec:initial_submission}
For our initial submission, we fit the model in Equation~\eqref{eq:model} using \textbf{flexBCF}.
We included averages of the beneficiary-level covariates at both pre-treatment time points as covariates for $\mu$ and similar averages at all four time points as covariates for $\tau.$
Although beneficiary-level covariates are time-invariant, practice composition generally changed over time as beneficiaries attrited from the study. 
Consequently, our conditioning on these post-treatment covariates can lead to biased estimation of the $\catt$ unless patient attrition is not impacted by treatment \citep{Rosenbaum1984_concomitant}.
We assumed that beneficiary attrition --- and therefore practice composition --- was not impacted by treatment.
We additionally included an estimated covariate balancing propensity score \citep{ImaiRatkovic2014} as a covariate for both $\mu$ and $\tau.$
Although we did not include outcomes measured in years 1 or 2 as covariates for $\mu$ or $\tau,$ we did include these measurements as covariates in our propensity score model.
We further placed a sparsity-inducing prior on the splitting variables for both the $\mu$ and $\tau$ tree ensembles.
This was in an attempt to discover which covariates drove heterogeneity in the prognostic function $\mu$ and treatment effect function $\tau.$

We submitted the posterior mean and 90\% credible interval of the requested ATT and subgroup SATT's, computed from the MCMC samples of evaluations $\tau(\bx,t)$ (and suitable averages thereof).
The credible intervals were formed by taking the 5\% and 95\% quantiles of the MCMC samples of the estimands. 
We note that these intervals, like all Bayesian credible intervals, are not theoretically guaranteed to have nominal frequentist coverage. 
After the Challenge concluded, the organizers provided us with the individual-level treatment effects, which we used to compute the RMSE and uncertainty interval coverage.

When fit to the beneficiary-level data, our initial submission achieved RMSE (uncertainty interval coverage) for the overall ATT between 9.2 (33\%) to 29.9 (82.5\%), with a median of 21.81 (49\%) and average of 20.54 (52.15\%) across all 17 DGP's.
For the SATT's, our RMSE's (coverages) ranged from 10.73 (26.5\%) to 39.72 (81\%). 

When fit to the practice-level data, our initial submission achieved RMSE (coverage) for the overall ATT between 11.11 (65\%) to 34.11 (99\%), with a median of 23.72 (87.5\%) and average of 23.04 (83.68\%) across all 17 DGP's.
Our RMSE's (coverages) ranged from 11.71 (61\%) to 44.29 (99.5\%) for individual SATT's. 

In sharp contrast to previous iterations of the ACIC Data Challenge, our BCF-based results were fairly middling compared to other submissions, especially in the presence of confounding; in fact, from what we could tell, our RMSE results tended to be around the median value.
Interestingly, when fit to the practice-level data, our uncertainty interval coverage improved substantially while the RMSE increased slightly. 
We suspect that the improvement in coverage is an artifact of sample size. 
Basically, the posterior distributions of the estimands seemed to be much more diffuse when conditioning on the practice-level data than on the beneficiary-level data.

\section{Re-analysis of Challenge Data}
\label{sec:results}
To assess the sensitivity of our initial submission results to our \textit{ad hoc} use of sparsity-inducing priors and choice of propensity score estimate, we re-analyzed the competition data using different propensity score estimates and using (or not) the sparsity-inducing prior on splitting variables for both $\mu$ and $\tau.$

We carried out two sensitivity analyses, one for the beneficiary-level data and one for the practice-level data.
For brevity, we report only the results for the beneficiary-level data here.
The findings for the practice-level data are qualitatively similar and are reported in Section~\ref{app:practice} of the Supplementary Materials.

For the beneficiary-level sensitivity analysis, we restricted our re-analysis to using sparsity-inducing priors in both ensembles or neither ensemble and did not consider the other possibilities (e.g.~sparsity in $\mu$ but not $\tau$ or \textit{vice versa}). 
In addition to the covariate balancing propensity score (hereafter \texttt{CBPS}), we considered running \textbf{flexBCF} with a propensity score estimated with gradient boosted trees (hereafter \texttt{GBM}).
We computed the two propensity score estimates using the \Rlang~packages \textbf{CBPS} \citep{cbps_package} and \textbf{GBM} \citep{gbm_package}. 
Note that \texttt{CBPS} assumes that the log-odds of treatment are linear in the covariates while \texttt{GBM} does not make any functional form assumptions. 

We refer to the model configuration used in our initial Challenge submission as \texttt{CBPS(S)} where the suffix \texttt{(S)} refers to the sparsity-inducing prior.
We use the suffix \texttt{(D)} to refer to the uniform prior on splitting variables. 
So for the beneficiary-level dataset, we considered four model configurations \texttt{CBPS(S)}, \texttt{CPBS(D)}, \texttt{GBM(S)}, and \texttt{GBM(D)}.

The 3,400 datasets consist of 200 replications of 17 different DGP's. 
For each combination of model configuration and dataset, we computed the posterior mean and a symmetric 90\% posterior credible interval using the posterior samples returned by \textbf{flexBCF} for the overall ATT and the SATT for each of the 12 pre-specified subgroups.
Using the true values provided by the Data Challenge organizers, we computed the squared difference between the posterior mean and the true value of the estimand.
We computed the RMSE as the square root of the average of these squared differences over the 200 replications of each DGP.
We further computed uncertainty interval coverage as the proportion of the 200 replications in which our credible interval contained the true value of the estimand. 

We report the RMSE and the uncertainty interval coverage within each of five categories of DGPs defined by the amount of confounding and effect heterogeneity.
The categories are: (i) no confounding but large effect heterogeneity; (ii) weak confounding with small effect heterogeneity; (iii) weak confounding with large effect heterogeneity; (iv) strong confounding with small effect heterogeneity; and (v) strong confounding with large effect heterogeneity.
The Data Challenge included exactly one DGP with no confounding and four DGPs in each of the other categories.

\subsection{Overall ATT}

\begin{figure}[ht]
    \centering
    \includegraphics[width=\textwidth]{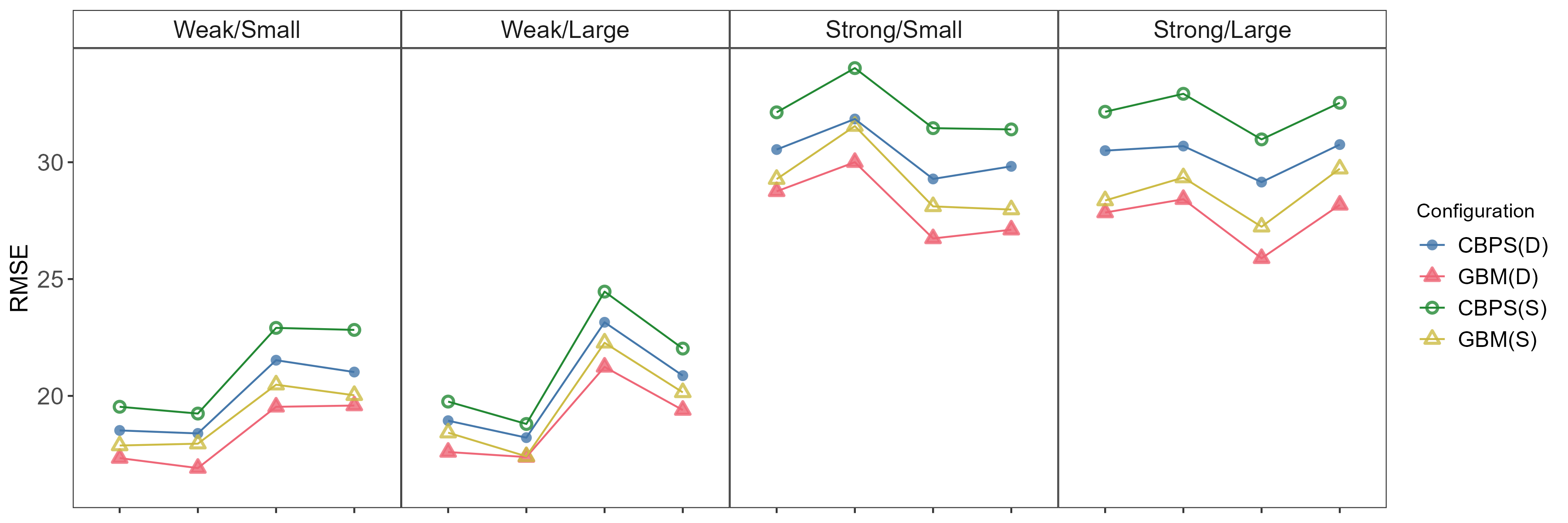}
    \caption{RMSE for overall ATT based on beneficiary-level data for the confounded DGPs. Dots represent average performance averaged across 200 replications of a DGP. Dots corresponding to the same model configuration are connected.}
    \label{fig:pt_overall_rmse}
\end{figure}

\begin{figure}[!ht]
    \centering
    \includegraphics[width=\textwidth]{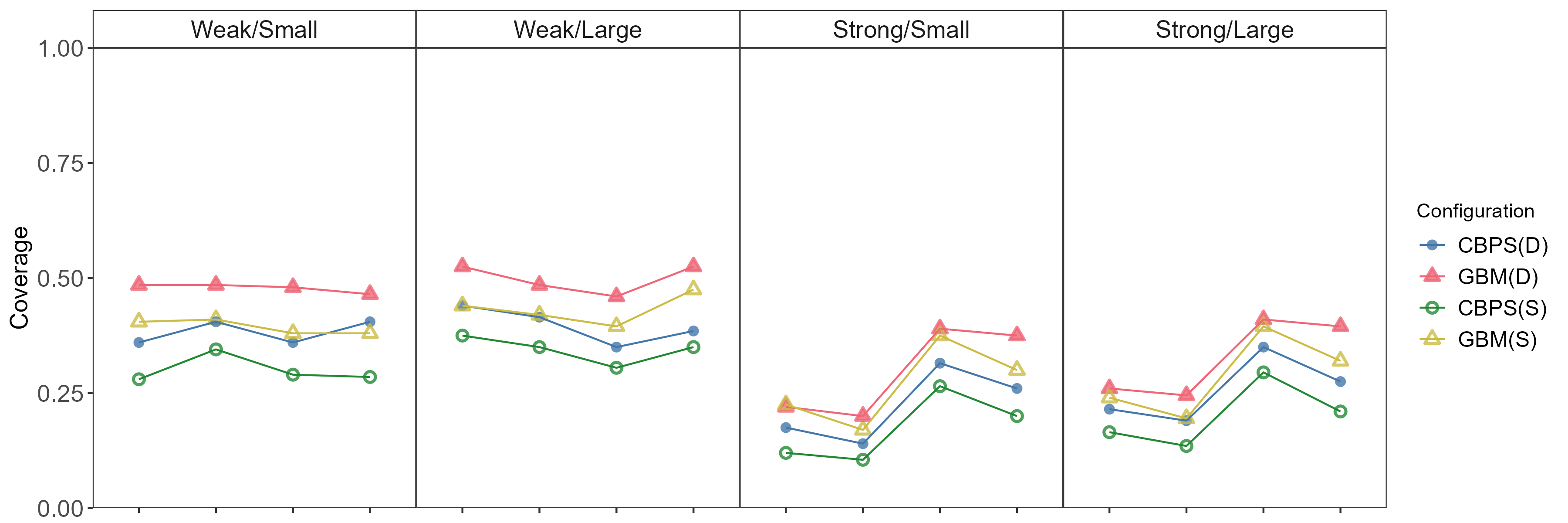}
    \caption{Uncertainty interval coverage for the overall ATT based on beneficiary-level data for the confounded DGPs. Dots represent performance averaged across 200 replications of a DGP. Dots corresponding to the same model configuration are connected.}
    \label{fig:pt_overall_cov}
\end{figure}

Generally speaking, we found that there were not massive differences between the four model configurations when it came to estimating the overall ATT.
Perhaps unsurprisingly, in the absence of confounding, each model configuration estimates the ATT quite accurately; 
the RMSE (coverage) in the unconfounded DGP was 9 (83.5\%) for \texttt{CBPS(S)}, 8.79 (84\%) for \texttt{CBPS(D)}, 9.28 (80.5\%) for \texttt{GBM(S)}, and 9.07 (85\%) for \texttt{GBM(D)}.
However, in the presence of confounding and large effect heterogeneity, we observed increased RMSE and decreased uncertainty interval coverage.
Figures~\ref{fig:pt_overall_rmse} and~\ref{fig:pt_overall_cov} shows the RMSE and coverage for the overall ATT for all model configurations and DGP's. 
Each point in the figures corresponds to a DGP.
We indicate the model configuration using the shape, fill, and color of each point, and we connect points corresponding to the same model configuration.

Interestingly, in the presence of confounding, the GBM-based configurations tended to achieve somewhat smaller RMSEs than the CBPS-based configurations across all DGP's.
We further observed that using sparsity-inducing priors consistently yielded slightly higher RMSE's. 
That said, the differences in performance are quite small.

Uncertainty interval coverage appeared more sensitive to our modeling choices.
For instance, using a CBPS-based propensity score often yielded somewhat worse coverage than using the more flexible GBM-based propensity score.
On further inspection, we found that the posterior intervals computed with the CBPS-based propensity score were shorter than the posterior intervals computing using the GBM-based propensity score; see Figure~\ref{fig:pt_intlenov} in Section~\ref{app:pt_intlen} of the Supplementary Materials for a comparison of interval lengths.

Across the board, each model configuration performed worse as the amount of confounding increased.
We suspect that this due to a certain amount of model misspecification.
Namely, we did not include pre-treatment outcomes or trends as covariates when fitting the model in Equation~\eqref{eq:model}.
It is conceivable that the failure to condition on these potential confounders led to the substantial bias (as evidenced by the large RMSE's).
The rather low uncertainty interval coverages are likely driven by the combination of the this bias and the extremely large sample size.
Said differently, the potential misspecification of our confounders and the large sample size likely led each of our BCF models to heavily concentrate around biased estimates. 

\subsection{Subgroup effects.}

\begin{figure}[ht]
    \centering
    \includegraphics[width=\textwidth]{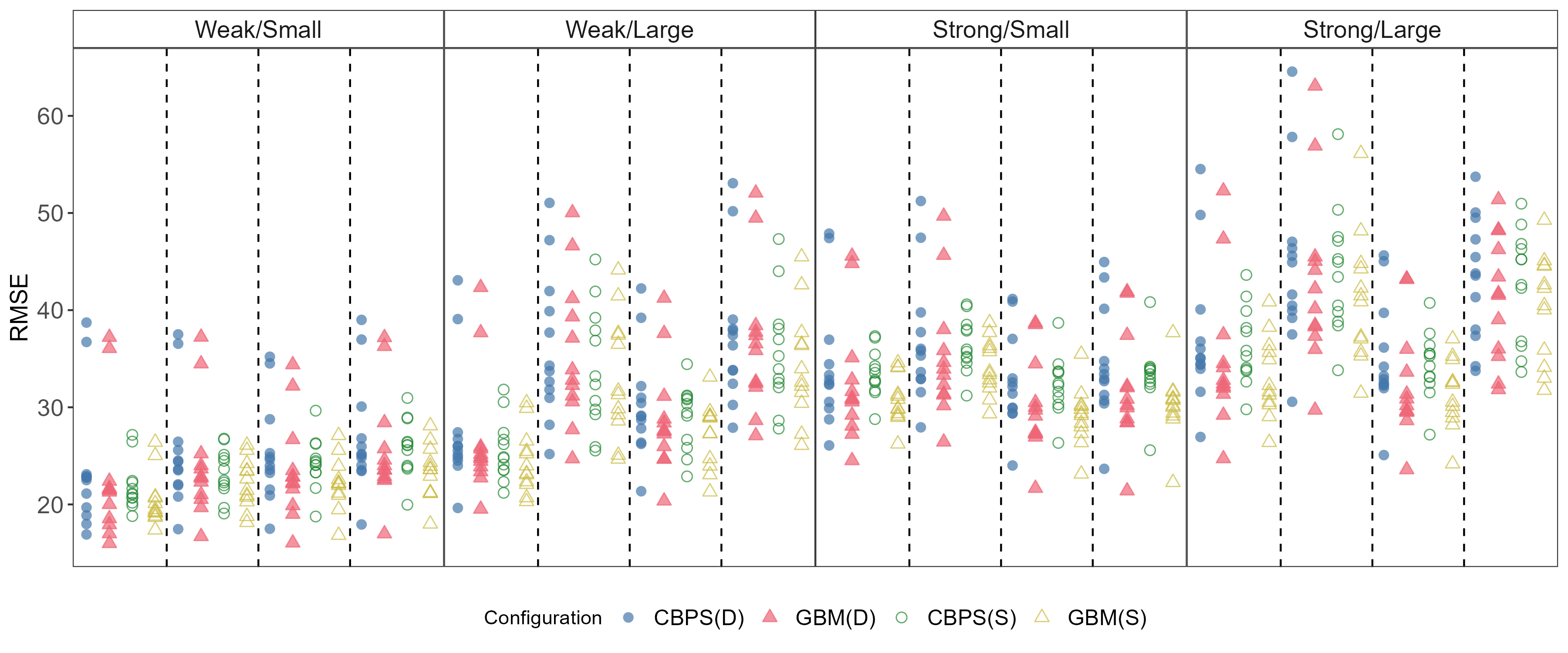}
    \caption{RMSE's for all 12 SATT's based on beneficiary-level data for the confounded DGPs. Dots represent average performance averaged across 200 replications of a DGP. All points between dashed vertical lines represent performance on the same DGP.}
    \label{fig:pt_rmse_comb}
\end{figure}

In the absence of confounding, the median RMSE (median coverage) across the 12 SATTs were 26.83 (49\%) for \texttt{CBPS(S)}, 30.04 (52\%) for \texttt{CBPS(D)}, 27.04 (52\%) for \texttt{GBM(S)}, and 30.35 (53\%) for \texttt{GBM(D)}.
Figures~\ref{fig:pt_rmse_comb} compares the RMSE's of each SATT for each model configuration across all categories of confounded DGP's.
We see that for all model configurations, RMSE increased alongside confounding and heterogeneity.
Additionally, we found that using the flexible GBM-based propensity score estimates yielded lower RMSE's than using CBPS-based estimates (triangles in Figure~\ref{fig:pt_rmse_comb} are shifted downwards compared to circles) and that sparsity-inducing priors produced slightly larger RMSE's (filled points shifted downwards compared to unfilled points). 

\begin{figure}[h]
    \centering
    \includegraphics[width=\textwidth]{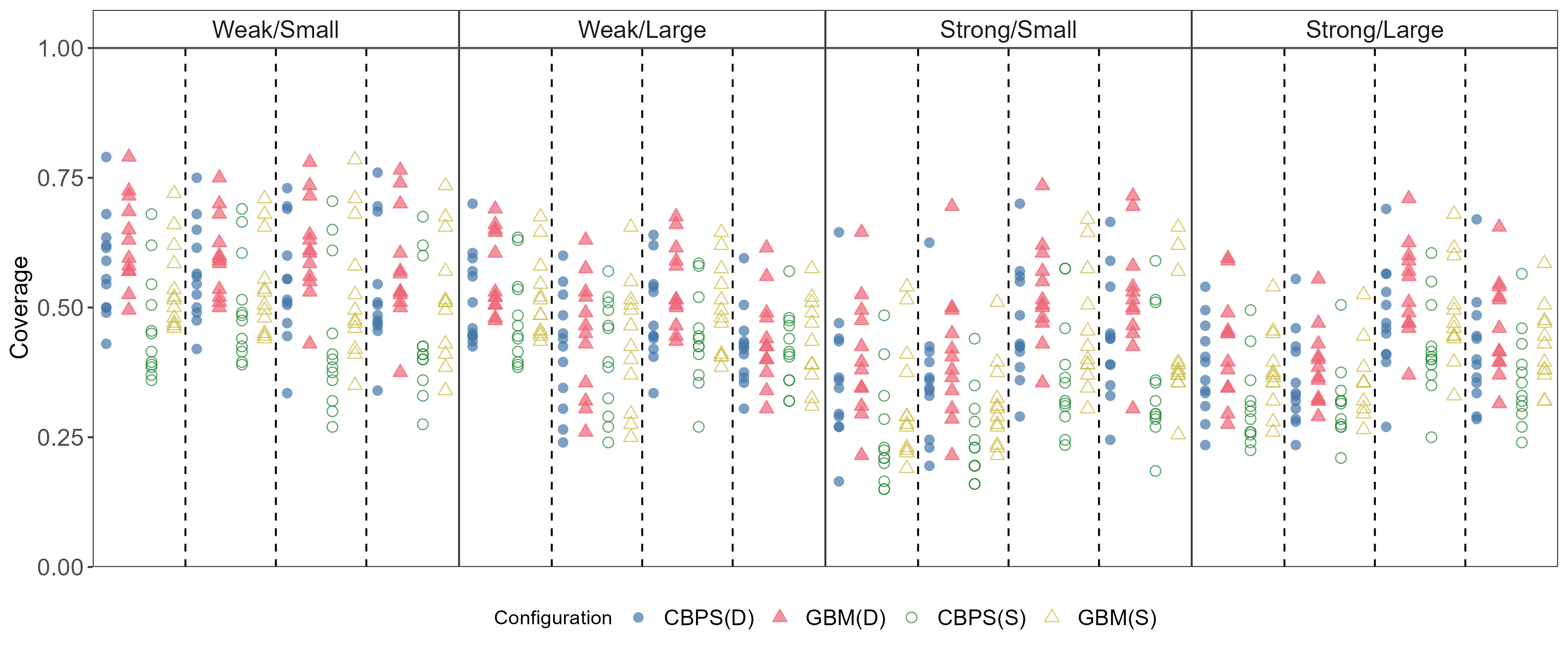}
    \caption{Uncertainty interval coverage for all 12 SATT's based on beneficiary-level data for the confounded DGP's. Dots represent average performance averaged across 200 replications of a DGP. All points between dashed vertical lines represent performance on the same DGP.}
        \label{fig:pt_cov_comb}
\end{figure}

Figure~\ref{fig:pt_cov_comb} compares the coverage of the uncertainty intervals across the 12 SATT's.
Generally speaking, the uncertainty interval coverage decreased as confounding and heterogeneity increased.
GBM-based models have very slightly higher coverage than the CBPS based models and using the sparsity-inducing priors consistently yielded lower coverage (un-filled points are shifted downwards compared to filled points in Figure~\ref{fig:pt_cov_comb}). 

\section{Discussion}
\label{sec:discussion}
We showed that this year's ACIC Data Challenge admitted a decomposition similar to the one used by \citet{Hahn2020}'s Bayesian casual forest model.
Because the \textbf{bcf} package could not scale to the size of the data in the competition, we implemented a faster and less resource-intensive version, which is available in the \textbf{flexBCF} package.
We investigated the sensitivity of our competition submission to two modeling choices: (i) the choice of propensity score method, and (ii) enforcing (or not) sparsity in the BART priors.
While we did observe some sensitivity (e.g.~flexible nonparametric propensity score estimates yielded longer uncertainty intervals with better coverage than parametric propensity score estimates), the differences in RMSE and coverage were not especially large.

As pointed out by a referee, the fact that the sparsity-inducing priors generally yielded worse performance may be because the Data Challenge organizers provided only those covariates that drove heterogeneity in the released datasets.
When all covariates are relevant, the sparsity-inducing prior may not have selected all covariates and instead may have inappropriately excluded some important drivers of heterogeneity.
Because we generally do not know all drivers of heterogeneity in practice, we do not anticipate that the results of our sensitivity analysis will generalize far beyond the Data Challenge.

While a small performance degradation can be expected as confounding increases, we were rather surprised by how poorly our BCF models performed in the presence of confounding.
We suspect that the poor performance stems from our decision to (i) directly model the observed outcomes using Equation~\eqref{eq:model} as opposed to trends and (ii) exclude pre-treatment outcome or trends as covariates in Equation~\eqref{eq:model}.
In a sense, we attribute the rather poor performance of our BCF models to model misspecification rather than any particular operating characteristic of BCF.

It is also possible that our specific results may be sensitive to other choices including, but not limited to, the inclusion (or exclusion) of certain covariates; the number of trees in the $\mu$ and $\tau$ ensembles; the choice of prior variance on the leaf node parameters in each tree; and the number of MCMC iterations and number of chains.
Due to time constraints, we could not exhaustively assess sensitivity along all these dimensions. 
However, we note that our package \textbf{flexBCF} renders such sensitivity analyses computationally feasible.
And while we do not anticipate our specific results to generalize beyond the Data Challenge, we do hope that \textbf{flexBCF} will facilitate the use of BCF on larger datasets. 


\section*{Acknowledgements}
We are grateful to Mariel Finucane, Jennifer Starling, and Dan Thal from Mathematica Policy Research for several helpful discussions following the Data Challenge. 

This research was performed using the computing resources and assistance of the UW--Madison Center For High Throughput Computing (CHTC) in the Department of Computer Sciences. 
The CHTC is supported by UW--Madison, the Advanced Computing Initiative, the Wisconsin Alumni Research Foundation, the Wisconsin Institutes for Discovery, and the National Science Foundation, and is an active member of the OSG Consortium, which is supported by the National Science Foundation and the U.S. Department of Energy's Office of Science.

Support for A.H.K. and S.K.D. was provided by the University of Wisconsin--Madison, Office of the Vice Chancellor for Research and Graduate Education with funding from the Wisconsin Alumni Research Foundation.

\vskip 0.2in
\bibliographystyle{apalike}
\bibliography{flexBCF_references}

\newpage
\appendix

\renewcommand{\theequation}{S\arabic{equation}}
\renewcommand{\thesection}{S\arabic{section}}  
\renewcommand{\thefigure}{S\arabic{figure}}  

{\begin{center} \Large \textbf{Supplementary Material} \end{center}}
\singlespacing

\section{Identification}
\label{app:identification}
\label{app:ident_proof_full}
Recall Assumption (A3), which asserts that there is a $0 < \delta < 1$ such that for all covariate vectors $\bx,$ $\delta < \P(Z_j = 1 \mid \bX_{ij} = \bx) < 1 - \delta.$
Marginalizing over $\bx,$ this implies that $\delta < \P(Z_{j} = 1) < 1- \delta$ so that the population consists of a non-trivial mixture of treated and untreated practices.


Let $t$ be a post-intervention time-point (i.e. $t \in \{3,4\}$) and let $s$ be a pre-intervention
time-point (i.e. $s \in \{1,2\}$). Then, for any $s$ and $t$ where $\delta < \P(Z_j = 1 \mid \bX_{ij} = \bx) < 1 - \delta$, we have the following: 
$$
    \begin{aligned}
        &\mathbb{E}\left[Y_{ijt} \mid Z_j=1, \bm{X}_{ij} = \bm{x}\right]-\mathbb{E}\left[Y_{ijt} \mid Z_j=0, \bm{X}_{ij} = \bm{x}\right]\\[15pt]
        &\, = \mathbb{E}\left[Y_{ijt}(1) \mid Z_j=  1, \bm{X}_{ij} = \bm{x}\right]-\mathbb{E}\left[Y_{ijt}(0) \mid Z_j=  0, \bm{X}_{ij} = \bm{x}\right] &(\text{by A1})\\[15pt] 
        &\, = \underbrace{\mathbb{E}\left[Y_{ijt}(1)-Y_{ijt}(0) \mid Z_j=  1, \bm{X}_{ij} = \bm{x}\right]}_{\operatorname{CATT}\left(\bm{x}, t\right)}\\
        &\,\quad+\mathbb{E}\left[Y_{ijt}(0) \mid Z_j=1, \bm{X}_{ij} = \bm{x}\right]-\mathbb{E}\left[Y_{ijt}(0) \mid Z_j=0, \bm{X}_{ij} = \bm{x}\right] \\[15pt]
        &\, =\operatorname{CATT}(\bm{x}, t)\\
        &\,\quad+\mathbb{E}\left[Y_{ijt}(0)-Y_{ijs}(0) \mid Z_j= 1, \bm{X}_{ij} = \bm{x}\right]-\mathbb{E}\left[Y_{ijt}(0)-Y_{ijs}(0) \mid Z_j=  0, \bm{X}_{ij} = \bm{x}\right] \\[10pt]
        &\,\quad+\mathbb{E}\left[Y_{ijs}(0) \mid Z_j=1, \bm{X}_{ij} = \bm{x}\right]-\mathbb{E}\left[Y_{ijs}(0) \mid Z_j=0, \bm{X}_{ij} = \bm{x}\right] \\[15pt]
        &\, = \operatorname{CATT}(\bm{x}, t)\\
        &\,\quad+ \underbrace{\mathbb{E}\left[Y_{ijt}(0)-Y_{ijs}(0) \mid \bm{X}_{ij} = \bm{x}\right]-\mathbb{E}\left[Y_{ijt}(0)-Y_{ijs}(0) \mid \bm{X}_{ij} = \bm{x}\right]}_{= 0} \\[10pt]
        &\,\quad+\mathbb{E}\left[Y_{ijs}(0) \mid Z_j=1, \bm{X}_{ij} = \bm{x}\right]-\mathbb{E}[Y_{ijs}(0) \mid Z_j=0, \bm{X}_{ij} = \bm{x}] &(\text{by A2}) \\[15pt]
        &\, = \operatorname{CATT}(\bm{x}, t)+\mathbb{E}\left[Y_{ijs}(0) \mid Z_j=  1, \bm{X}_{ij} = \bm{x}\right]-\mathbb{E}\left[Y_{ijs}(0) \mid Z_j=0, \bm{X}_{ij} = \bm{x}\right] \\[10pt]
        &\, = \operatorname{CATT}(\bm{x}, t)+\mathbb{E}\left[Y_{ijs}|Z_j= 1, \bm{X}_{ij} = \bm{x}\right] - \mathbb{E}\left[Y_{ijs} \mid Z_j=0, \bm{X}_{ij} = \bm{x}\right] &(\text{by A1})
    \end{aligned}
$$
Rearranging we can express the estimand of interest in terms of expectations of observables as follows:
\begin{align*}
    \begin{split}
    \catt(\bx, t) \ =\ & 
        \mathbb{E}\left[Y_{ijt} \mid Z_j = 1, \bX_{ij} = \bx \right] - \mathbb{E}\left[Y_{ijt} \mid Z_j = 0, \bX_{ij} = \bx \right]\\ 
        &- \Big\{ 
        \mathbb{E}\left[Y_{ijs} \mid Z_j = 1, \bX_{ij} = \bx \right] - \mathbb{E}\left[Y_{ijs} \mid Z_j = 0, \bX_{ij } = \bx \right]\Big\},
    \end{split}
\end{align*}
which is the same as Equation~\eqref{eq:main_identification}. 


\section{Practice-level sensitivity analysis}
\label{app:practice}
In Section~\ref{sec:results} of the main text, we reported the results of a sensitivity analysis run on the beneficiary-level data.
Here, we repeat that sensitivity analysis using the practice-level data.
Because the practice-level data was so much smaller, we were able to investigate the sensitivity to more choices of the propensity score model.
In addition to the \texttt{CBPS} and \texttt{GBM}, we considered propensity scores estimated with BART (hereafter \texttt{BART}) and $L_{1}$-regularized logistic regression (hereafter \texttt{LASSO}).
We computed the four propensity score estimates using the \Rlang~packages \textbf{CBPS} \citep{cbps_package}, \textbf{BART} \citep{bart_package}, \textbf{glmnet} \citep{glmnet_package}, and \textbf{GBM} \citep{gbm_package}. 
Note that \texttt{CBPS} and \texttt{LASSO} assume that the log-odds of treatment are linear in the covariates $\bx$ while \texttt{BART} and \texttt{GBM} do not make any functional form assumptions about the log-odds of treatment.
Accordingly, we refer to \texttt{CBPS} and \texttt{LASSO} as parametric and \texttt{BART} and \texttt{GBM} as non-parametric propensity score estimators.
We will refer to the model used in our initial Challenge submission as \texttt{CBPS(S)} where the suffix \texttt{(S)} refers to the sparsity-inducing prior.
We will use the suffix \texttt{(D)} to refer to the uniform prior on splitting variables. 

\subsection{Overall ATT}

\begin{figure}[ht]
    \centering 
    \includegraphics[width = \textwidth]{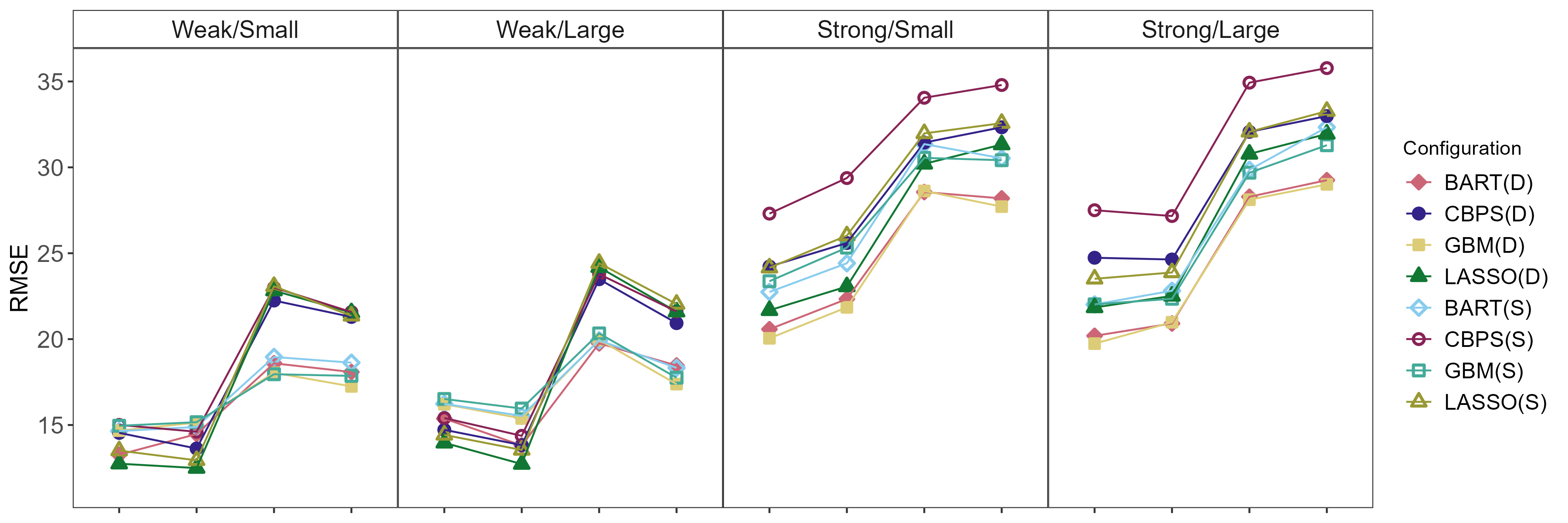}
    \caption{RMSE for overall ATT based on practice-level data for the confounded DGP's. Dots represent average performance averaged across 200 replications of a DGP. Dots corresponding to the same model configuration are connected.}
    \label{fig:rmseov}
\end{figure}

For the uncofounded DGP, the median RMSEs for the overall ATT ranged from 9.9 (\texttt{LASSO(D)}) to 13.7 (\texttt{GBM(D)}) while the uncertainty interval coverages ranged from 83\% (\texttt{GBM(D)}) to 94.5\% (\texttt{LASSO(D)}).
Figure~\ref{fig:rmseov} compares the RMSEs of all methods for the overall ATT for each confounded DGP broken down by category.
In general, as the amount of confounding and treatment effect heterogeneity increased, the RMSE of all methods increased.
While we did not observe substantial differences between the model configurations in presence of weak confounding, the \texttt{CBPS(S)} configuration did worse than all other models when there was strong confounding.

\begin{figure}[ht]
    \centering 
    \includegraphics[width = \textwidth]{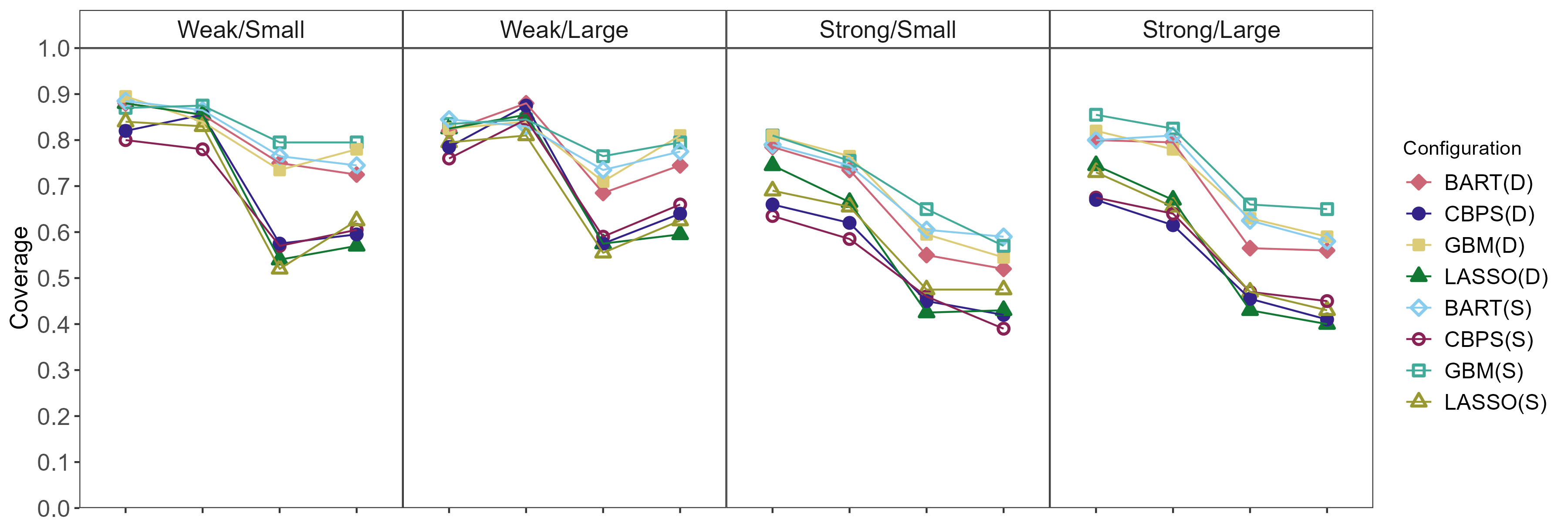}
    \caption{Coverage of the uncertainty intervals for the overall ATT for the DGP categories with confounding using practice-level data.
    }
    \label{fig:intscov}
\end{figure}

Interestingly, in the absence of confounding, using a nonparametric propensity score estimate (i.e.~\texttt{BART} or \texttt{GBM}) yielded slightly larger RMSE than using parametric estimates (i.e.~\texttt{CBPS} or \texttt{LASSO}).
However, in the presence of even weak confounding, the nonparametric propensity score estimates fared slightly better.
Figure~\ref{fig:intscov} shows the coverage for the uncertainty intervals for the ATT obtained from the practice-level model for each confounded DGP.

\begin{figure}[ht]
    \centering
    \includegraphics[width = \textwidth]{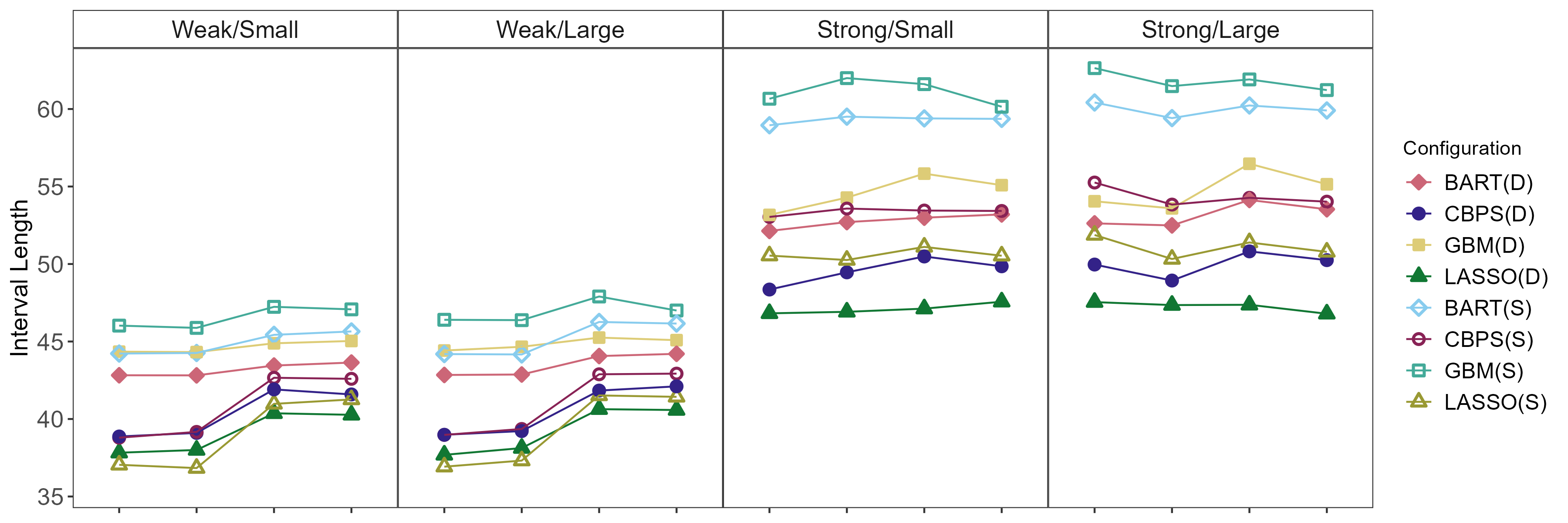}
    \caption{Length of the uncertainty intervals for the overall ATT for the DGP categories with confounding using practice-level data. 
    }
    \label{fig:intlenov}
\end{figure}

In the presence of confounding, we further found that fitting BCF with flexible propensity score estimate (i.e.~\texttt{BART} or \texttt{GBM}) yielded wider uncertainty intervals with much better frequentist coverage than parametric propensity score estimates (i.e.~\texttt{CBPS} or \texttt{LASSO}); see Figure~\ref{fig:intlenov}.

\subsection{Subgroup effects}

\begin{figure}[ht]
    \centering
    \includegraphics[width=\textwidth]{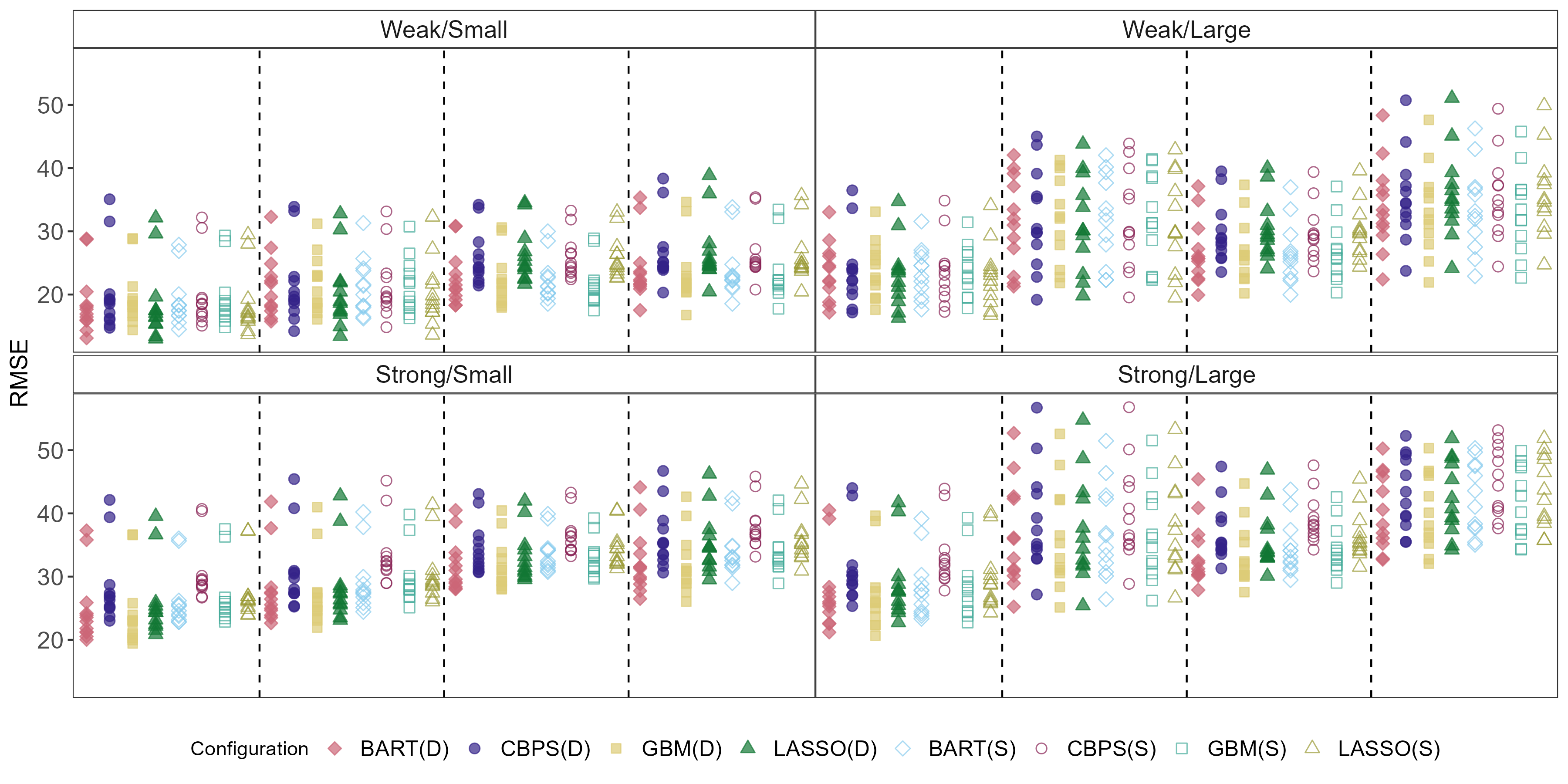}
    \caption{RMSE's for all 12 SATT's based on practice-level data for the confounded DGPs. Dots represent average performance averaged across 200 replications of a DGP. All points between dashed vertical lines represent performance on the same DGP.}
    \label{fig:prac_rmse_sub}
\end{figure}

We similarly compared the estimation performance and uncertainty quantification on the eight models for each subgroup effect.
For the unconfounded DGP, the RMSEs for the subgroup ATTs ranged from 17.32 to 39.34 across subgroup estimands and model configurations.
Figure~\ref{fig:prac_rmse_sub} compares the RMSE's of each SATT for each model configuration across all categories of confounded DGP's.
We see that for all model configurations, RMSE increased alongside confounding and heterogeneity.
Similar to the overall ATT, we did not observe substantial differences in the quality of estimated subgroup effects across the different methods.

\begin{figure}[ht]
    \centering
    \includegraphics[width=\textwidth]{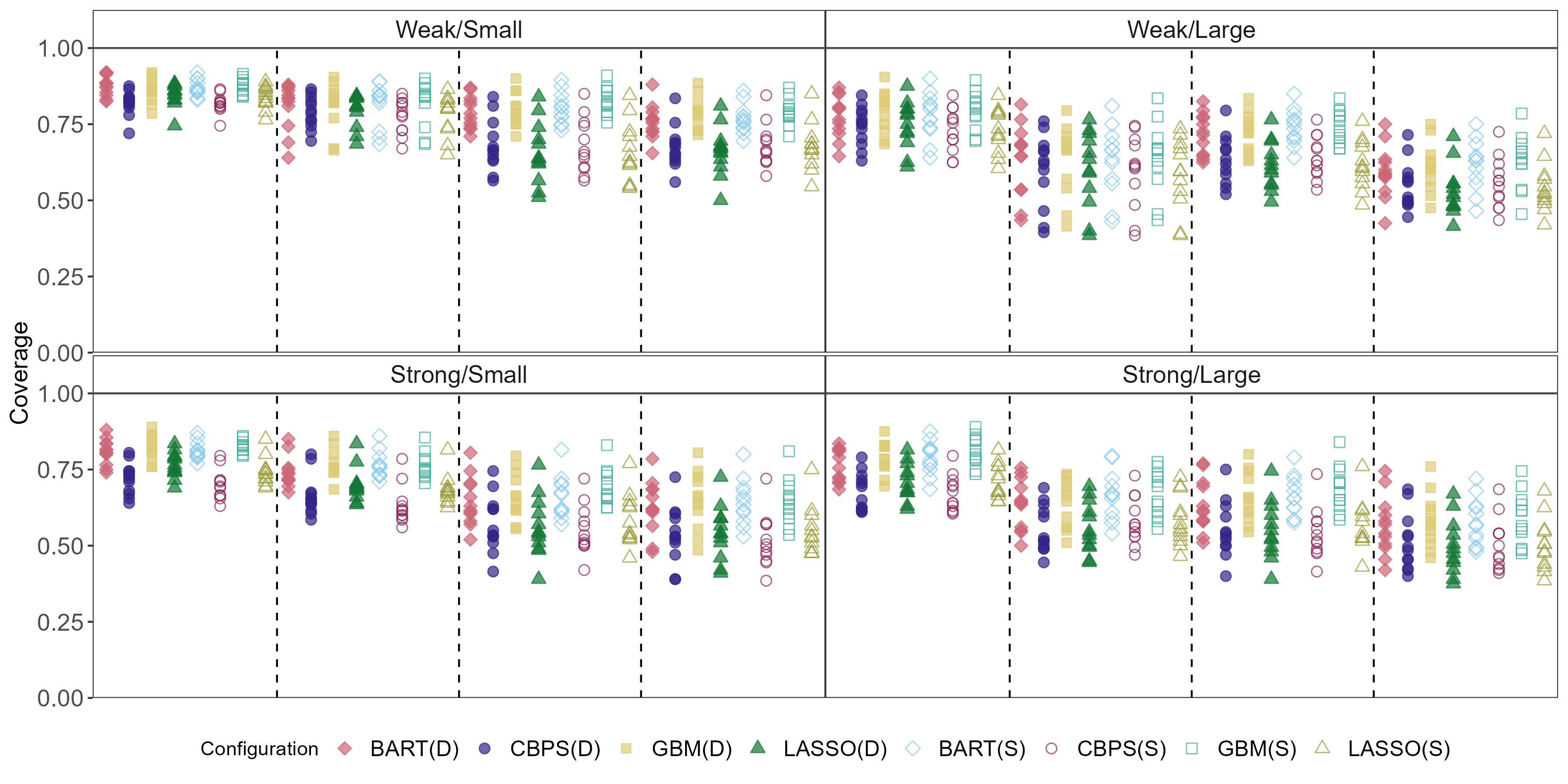}
    \caption{Uncertainty interval coverage for all 12 SATT's based on practice-level data for the confounded DGP's. Dots represent average performance averaged across 200 replications of a DGP. All points between dashed vertical lines represent performance on the same DGP.}
        \label{fig:prac_cov_subg}
\end{figure}

Next, we compare the coverage of the 90\% posterior credible intervals.
For the unconfounded DGP, the coverage for the subgroup ATTs ranged from 37.5\% to 83.5\% across subgroup estimands and model configurations.
Figure~\ref{fig:prac_cov_subg} compares the coverage of the uncertainty intervals for each of the 12 subgroup ATTs for every confounded DGP and model configuration.
Observe that the non-parametric propensity score estimators yielded posterior credible intervals with higher frequentist coverage than parametric propensity score estimators.
This may be attributed to the fact that the model configurations with non-parametric propensity scores yield slightly longer uncertainty intervals as can be seen in Figure~\ref{fig:prac_intlen_subg}.

\begin{figure}[ht]
    \centering
    \includegraphics[width=\textwidth]{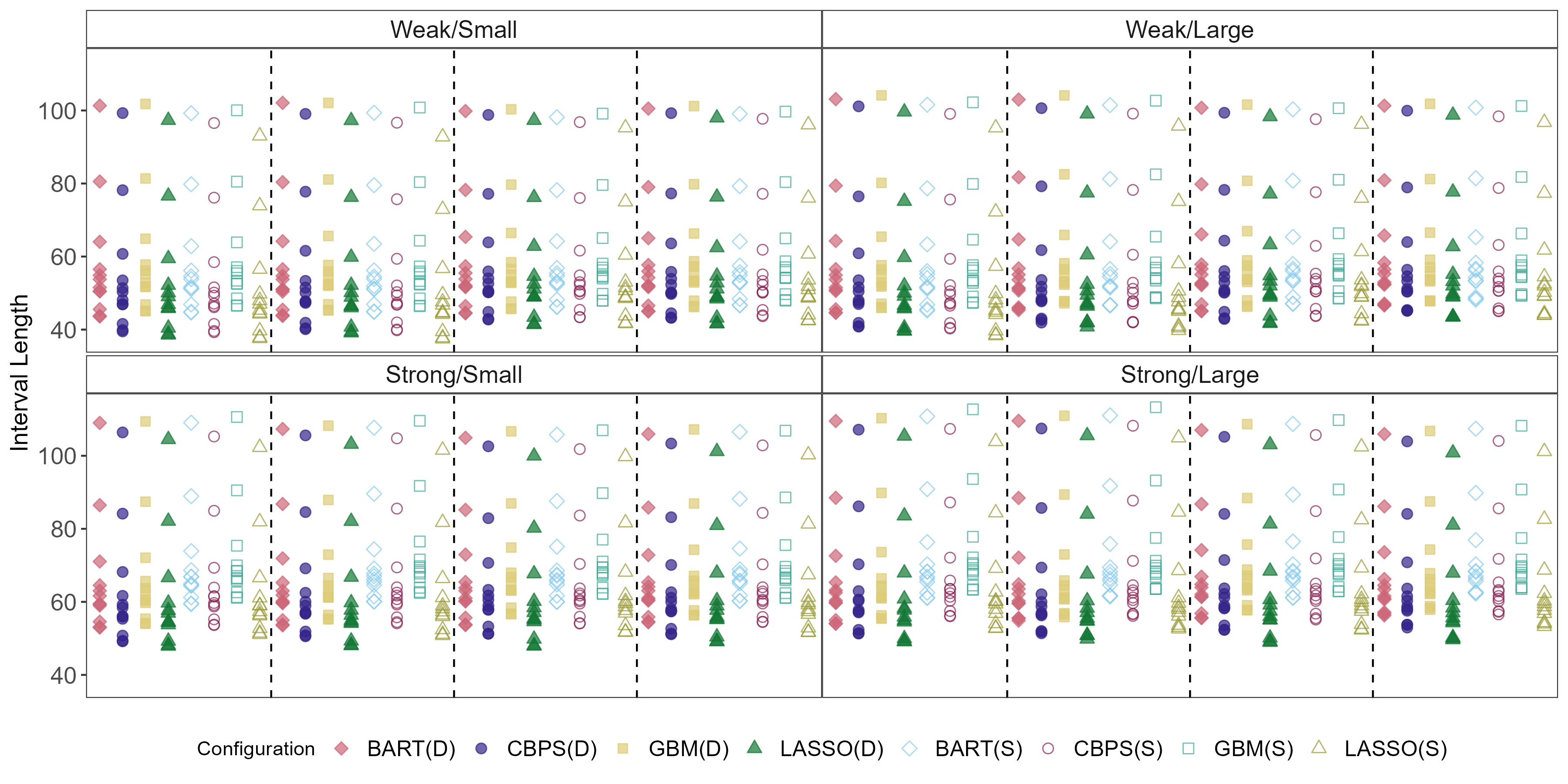}
    \caption{Length of the uncertainty intervals for all 12 SATT's based on practice-level data for the confounded DGP's. Dots represent average performance averaged across 200 replications of a DGP. All points between dashed vertical lines represent performance on the same DGP.}
        \label{fig:prac_intlen_subg}
\end{figure}
To summarize, the practice-level analyses do not indicate a clear preference for any of the eight models we compared in terms of point estimation.
However, we did observe that BCF fit using flexible propensity score estimates produced better-calibrated uncertainty intervals than BCF fit with parametric propensity score estimates.
Finally, the use of sparsity-inducing priors did not seem to affect performance much at all.
As pointed out by a referee, the apparent lack of sensitivity to the use of sparsity-inducing priors may stem from the fact that the Data Challenge organizers may have only included covariates that drove heterogeneity in the dataset.
Without access to the original DGP, however, it is impossible to verify whether all provided covariate actually drove heterogeneity in the treatment effects.

\section{Additional figures}
\label{app:pt_intlen}
\begin{figure}[H]
    \centering
    \includegraphics[width = \textwidth]{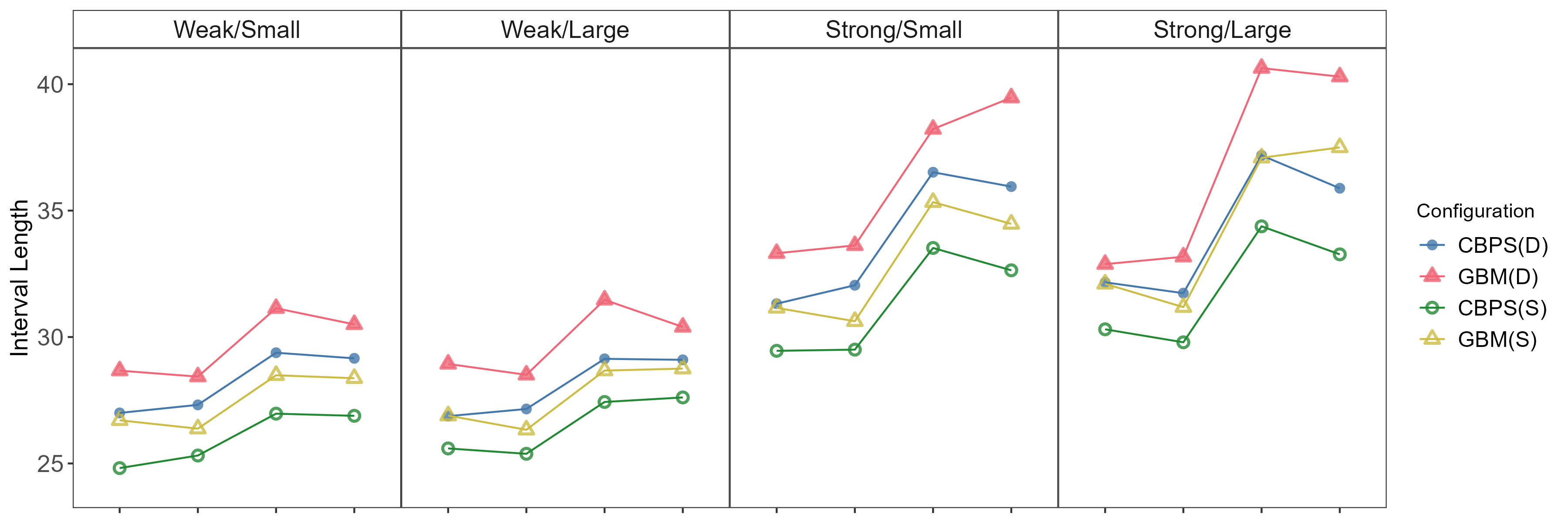}
    \caption{Length of the uncertainty intervals for the overall ATT for the DGP categories with confounding using beneficiary-level data. 
    }
    \label{fig:pt_intlenov}
\end{figure}

\begin{figure}[H]
    \centering
    \includegraphics[width=\textwidth]{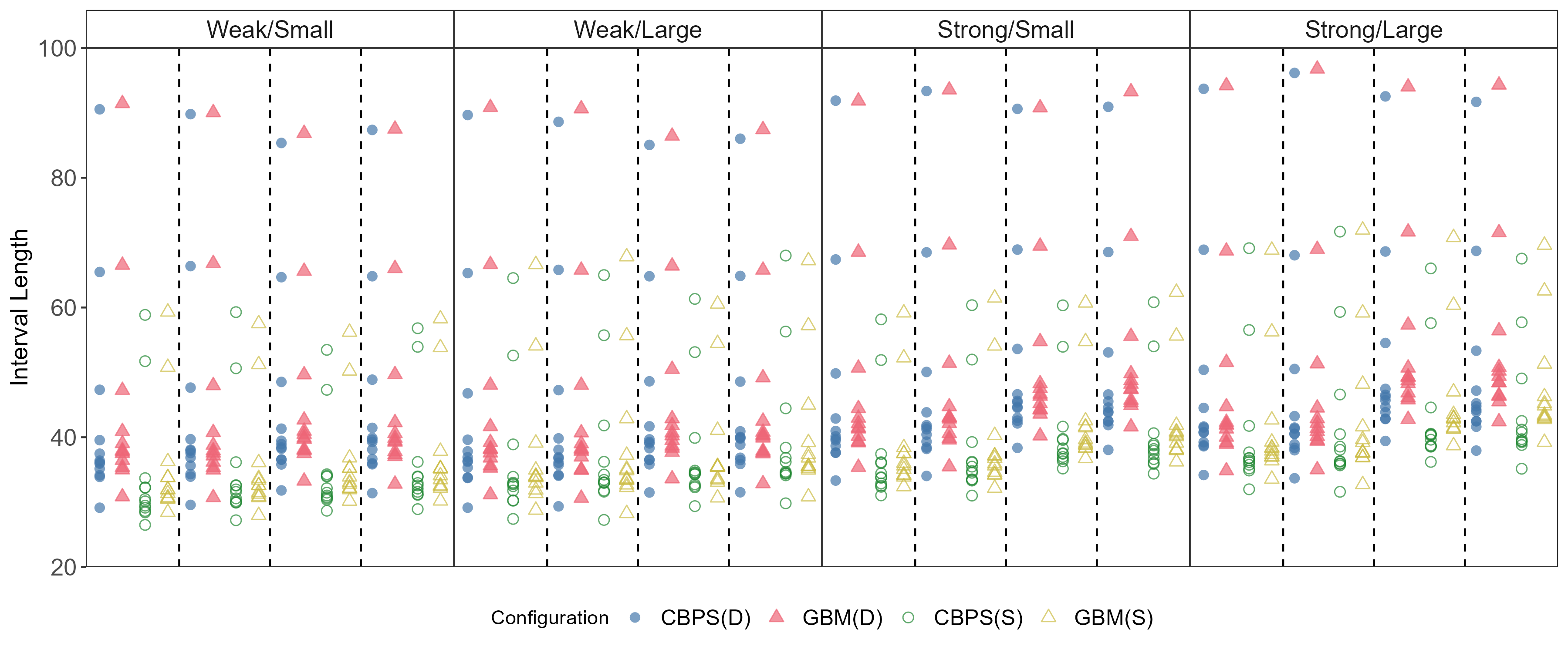}
    \caption{Length of the uncertainty intervals for all 12 SATT's based on beneficiary-level data for the confounded DGP's. Dots represent average performance averaged across 200 replications of a DGP. All points between dashed vertical lines represent performance on the same DGP.}
        \label{fig:prac_intlen_subg}
\end{figure}

\end{document}